\documentclass[preprint,showpacs,preprintnumbers,amsmath,amssymb,prb,superscriptaddress]{revtex4-1}

\usepackage[usenames,dvipsnames]{color}
\usepackage{graphicx}
\usepackage{dcolumn}
\usepackage{simplewick}
\usepackage{bm}
\usepackage{amsmath}
\usepackage{float}
\usepackage{threeparttable, tablefootnote}
\usepackage{hyperref}
\usepackage{natbib}
\usepackage{hyperref}% add hypertext capabilities
\hypersetup{colorlinks=true, urlcolor=blue, citecolor=blue}
\usepackage{multirow}
\usepackage{relsize}

\newcommand{\cgx}{{\rm CsGeX$_3$}}
\newcommand{\cgb}{{\rm CsGeBr$_3$}}
\newcommand{\cgc}{{\rm CsGeCl$_3$}}
\newcommand{\cgi}{{\rm CsGeI$_3$}}
\begin{document}

%\title{DFT-based classical potentials for ferroelectric perovskites with lone-pair: the case of  CsGeX$_3$ (X=Cl, Br, I)}

\title{DFT-based insight into finite-temperature properties of  ferroelectric perovskites with lone-pair: the case of  CsGeX$_3$ (X=Cl, Br, I)}

\author{Ravi Kashikar}
\email{ravik@usf.edu}
\affiliation{Department of Basic Sciences, Institute of Infrastructure, Technology, Research and Management(IITRAM), Ahmedabad, Gujarat 380026, India}
\author{S. Lisenkov}
\author{I. Ponomareva}
\email{iponomar@usf.edu}
\affiliation{Department of Physics, University of South Florida, Tampa, Florida 33620, USA}

\date{\today}

\begin{abstract}

Ferroelectrics remain in the focus of scientific attention for decades owing to their fundamental and practical appeal. Recently, ferroelectricity has been demonstrated in semiconducting halide perovskites\cite{CGX_Ferro}, offering both a rare combination of ferroelectricity and semiconductivity in the same material and a possible alternative to the prevailing perovskite oxide ferroelectrics. We propose a route to simulating such materials at finite temperatures capable of reproducing key experimental and first-principle data, such as Curie temperature, phase transition sequence, spontaneous polarization, and soft mode frequencies. The key methodological finding is the superior performance of hybrid exchange correlation functionals in parametrization of effective Hamiltonians for ferroelectrics with lone pair. The parametrization for effective Hamiltonians for \cgx\, (X=Cl, Br, I)  is reported. The application of methodology to study polarization reversal in \cgx\, allows for the development of a ``minimalistic"  model for polarization reversal in ferroelectrics that provides an insight into the mechanisms of polarization reversal and its key features, such as the relationship between the coercive field, temperature, and AC field frequency. Importantly, the model reveals the origin of the well-known and ever-puzzling overestimation of coercive fields in computations. Furthermore, we report a variety of finite-temperature properties of \cgx\,  ferroelectrics, such as dielectric susceptibility, pyroelectric coefficients, and energy storage density, which reveal that these halide perovskites possess properties comparable to their oxide counterparts. We believe that our work provides significant methodological advancements, deepens fundamental understanding of ferroelectrics, and reveals the potential of halide perovskite ferroelectrics.

\end{abstract}
\maketitle

\section{Introduction}

Ferroelectrics are materials that exhibit spontaneous polarization that is reversible by an external electric field. Typically, they undergo structural phase transitions at finite temperatures, which result in the onset of spontaneous polarization\cite{lines2001principles}. The spontaneous polarization is a function of both temperature and strain, which results in the materials ability to convert between electrical, mechanical and thermal energies, giving origin to numerous applications of ferroelectrics such as memory devices, sensors, energy harvesting, and nonlinear optical devices\cite{scott2007applications, devices1, devices2}. A good ferroelectric typically has large spontaneous polarization and high Curie temperatures. Many belong to the perovskite oxide family with BaTiO$_3$, BiFeO$_3$, and PbTiO$_3$ being the famous examples \cite{lines2001principles}. Such oxides are typically wide-bandgap insulators. Recently, ferroelectricity has been discovered in the family of semiconducting halide perovskites, \cgx\ (X=Cl, Br, I)\cite{CGX_Ferro}. The possibility to combine ferroelectricity and semiconducting nature is very attractive, as it could open a way to non-volatile and tunable spintronics applications. For example, this family has already been found to exhibit Rashba effects \cite{Zunger-Ferro, Popoola-JPCC} that are tunable by external electric field \cite{Popoola-JPCC}. Moreover, the possibility to convert charge-to-spin via Rashba-Eldestein effect has recently been computationally discovered in the same family \cite{RE-Popoola}. However, the same semiconducting nature can become a challenge,  for example, when ferroelectric properties are characterized experimentally. For example, significant conductivity of such samples complicates measuring spontaneous polarization\cite{semi_ferro}.  We are not aware of any reports that assess the temperature evolution of spontaneous polarization in these materials. Likewise, the temperature evolution of ferroelectric hysteresis loops also remains unknown. 

Computer simulations are an attractive alternative for understanding ferroelectricity and associated phase transitions in such materials. To that end, the effective Hamiltonian technique \cite{Effective_H0, Effective_H1} is arguably the most widely used computational tool for finite-temperature modeling of ferroelectrics. It is based on retaining degrees of freedom most relevant for ferroelectric phase transitions and writing the Hamiltonian in terms of the expansion into certain symmetry invariants of these degrees of freedom. The terms of the effective Hamiltonian that describe the energy contributions associated with different interactions are then parametrized with the help of first-principles simulations, such as density functional theory (DFT) simulations. The effective Hamiltonian can be used within the frameworks of Monte Carlo or Molecular Dynamics (MD) simulations to access finite-temperature properties of ferroelectrics \cite{Effective_H1, Krakauer01021998,PhysRevB.77.012102}. Effective Hamiltonian techniques have a long history of successful applications.  Examples include phase transitions in ferroelectrics \cite{Effective_H1,PTO_phase}, antiferroelectrics \cite{PhysRevB.91.134112}, multiferroics \cite{BFO}, ferroelectric solid solutions \cite{PZT}, ferroelectrics with defects \cite{PhysRevB.93.134101},  the dynamics of phase transitions \cite{PhysRevB.77.012102}, prediction of nanodomain phases in ferroelectric nanostructures \cite{NaumovIvanI2004Upti, PhysRevLett.93.196104, PhysRevLett.132.026902, PhysRevLett.132.136801}, the flexoelectric effect \cite{PhysRevB.85.104101}, and electro-, elasto-, piezo-, and multicaloric effects \cite{PhysRevB.78.052103, PhysRevB.86.104103, PhysRevB.87.224101, PhysRevB.93.064108}.  
 Recently, the effective Hamiltonian has been developed for \cgb \cite{Ravi_PRB} and successfully applied to predict the existence of dynamic antipolar phase above the Curie temperature of this material, compute temperature-stress and temperature-strain phase diagrams \cite{Townsend}, and predict scaling down properties of \cgb\,  films\cite{Ravi-NL}.  However, one shortcoming of the effective Hamiltonian developed for \cgb\, is its underestimation of the Curie point. The inability of the effective Hamiltonian techniques to accurately reproduce experimental Curie points is well-known issue, which is believed to originate from the performance of the exchange correlation functional\cite{XC-3, XC-2, XC-4, XC-1, XC-5, Maggie_PRM}. As a result, the effective Hamiltonian that does not accurately capture the Curie temperature can be used to obtain qualitative rather than quantitative predictions. However, in some instances, qualitative predictions are not sufficient. One such example is the electrocaloric effect, which describes the adiabatic change in temperature under application of electric field. According to the Maxwell relation, the electrocaloric change in temperature can be estimated as $ \Delta T = -\int_{E_1}^{E_2} \frac{T}{C_E} \left( \frac{\partial P}{\partial T} \right)_E dE$, where $P$, $T$ and $E$ are the polarization, temperature and applied electric field, while $C_E$ is the heat capacity under constant electric field.  The derivative $\left( \frac{\partial P}{\partial T} \right)_E$, and, consequently, the electrocaloric $\Delta T$, is maximized in the vicinity of the phase transition so we can estimate the largest electrocaloric change in temperature as -$\Delta T \approx \frac{T_C}{C_E} \left( \frac{\partial P}{\partial T} \right)_{E_1,T_C} (E_2-E_1)$. Therefore, inaccuracy in $T_C$ is detrimental for the predictions of this response. In addition, there is no parametrization available for \cgc\, and \cgi.  
Therefore the aims of this work are: (i) to propose a  strategy  for the effective Hamiltonian parametrization capable of reproducing experimental Curie temperatures in \cgx; (ii) to develop parametrization for \cgx\, using the proposed strategy; (iii) to utilize the effective Hamiltonians to predict finite-temperature properties of \cgx, such as hysteresis loops, susceptibility, pyroelectric response and energy storage density; (iv) to propose a basic  but powerful model for the polarization reversal in ferroelectrics that provides insight into the mechanisms of polarization reversal, its intrinsic time scales, and reveals the origin of the well-known coercive field overestimation  in simulations.  The last aim is especially significant  in the light of a renewed interests to polarization reversal as a mean to realize neuromorphic computing \cite{https://doi.org/10.1002/adma.201905764,LiangLei2025Ckop}
It should be noted that in the recent years Machine Learned potentials are emerging as powerful simulations tools for ferroelectrics at finite temperatures \cite{PhysRevB.107.014101,PhysRevB.108.L180104,zhang2024}. Nevertheless, as argued below, we believe that they may not be able to provide a viable route for accurate parametrization of \cgx\, just yet.

\section{Computational Methodology}
\cgx\, are known to undergo a single phase transition from Pm$\Bar{3}$m to R3m phase at temperatures 428 K, 511 K, 548 K for X$=$Cl, Br, I, respectively \cite{thiele1987kristallstrukturen}. Experimentally, the polarization values have been reported 12-15 $\mu$C/cm$^2$ in \cgb\ and 20 $\mu$C/cm$^2$ in \cgi\ \cite{CGX_Ferro}. Computationally, the polarization 24.2 $\mu$C/cm$^2$, 22.2 $\mu$C/cm$^2$, and 20.2 $\mu$C/cm$^2$ in \cgc\, \cgb\ and \cgi\, respectively  have been reported. We aim to develop an effective Hamiltonian that reproduces both experimental Curie temperatures and DFT zero Kelvin polarization. The reason for the latter is that, first of all, polarization are less reliable from experiment because of the aforementioned issues and are also scarce. Second of all, polarization from DFT are, on the contrary, reliable and show little to no dependence on the exchange-correlation functional\cite{Maggie_PRM}. Here we follow the strategy of Refs.\onlinecite{Effective_H1,Ravi_PRB} and retain only ferroelectric soft mode and strain as the relevant distortions for ferroelectric phase transition. 
Consequently, the degrees of freedom for the effective Hamiltonian are  local modes, which are proportional to the electric dipole moment of perovskite five atom unitc cell, $\{\mathbf{u}_i\}$, and local strain tensors, $\{\eta_l\}$, (in Voigt notations) that  describe unit cell deformation \cite{Effective_H1}. The local modes describe the ferroelectric soft mode. The Hamiltonian is given by \cite{Effective_H1}:
\begin{equation}
    E_{tot} = E_{self}(\{\mathbf{u}_i\})+E_{dpl}(\{\mathbf{u_i}\})+E_{short}(\{\textbf{u}_i\})+E_{elas}(\{\eta_l\})+E_{int}(\{\textbf{u}_i\}, \{\eta_l\})
\label{heff}
\end{equation}
where the terms describe local mode self energy (harmonic and anharmonic contributions), a long-range dipole-dipole interaction, a short-range interaction between local modes, an elastic energy, and the interaction between the local modes and strains, respectively.
In Ref.\onlinecite{Ravi_PRB}, the effective Hamiltonian has been parameterized for \cgb\, using PBE and r$^2$SCAN exchange-correlation functionals. Both functionals predicted the correct phase transition sequence from the cubic (paraelectric) phase to the rhombohedral (ferroelectric) phase. However, the Curie temperatures (270~K and 350~K with PBE and r$^2$SCAN, respectively) underestimated the experimental one of 511~K. It was noted in Ref.\onlinecite{Townsend} that the hybrid HSE exchange-correlation functional may help overcome the issue, based on its prediction of larger energy differences between cubic and rhombohedral phases, $\Delta U$, of the material.  Indeed,   it is presently, believed that ferroelectric distortion of \cgx\, originates from second-order Jahn-Teller instability, which distorts GeX$_6$ octahedra to accommodate Ge lone pair \cite{doi:10.1021/acs.jpcc.2c06818, doi:10.1021/ic970659e}. However, such lone pair is traditionally not well described by non-hybrid exchange correlation functionals due to their intrinsic tendency to delocalize charge density \cite{Zunger_lone-pair}. Therefore, using a hybrid exchange correlation functional may offer a way to develop more accurate parametrization. This work aims to provide a comprehensive answer to this question.

\begin{figure}
\centering
\includegraphics[width=1\textwidth]{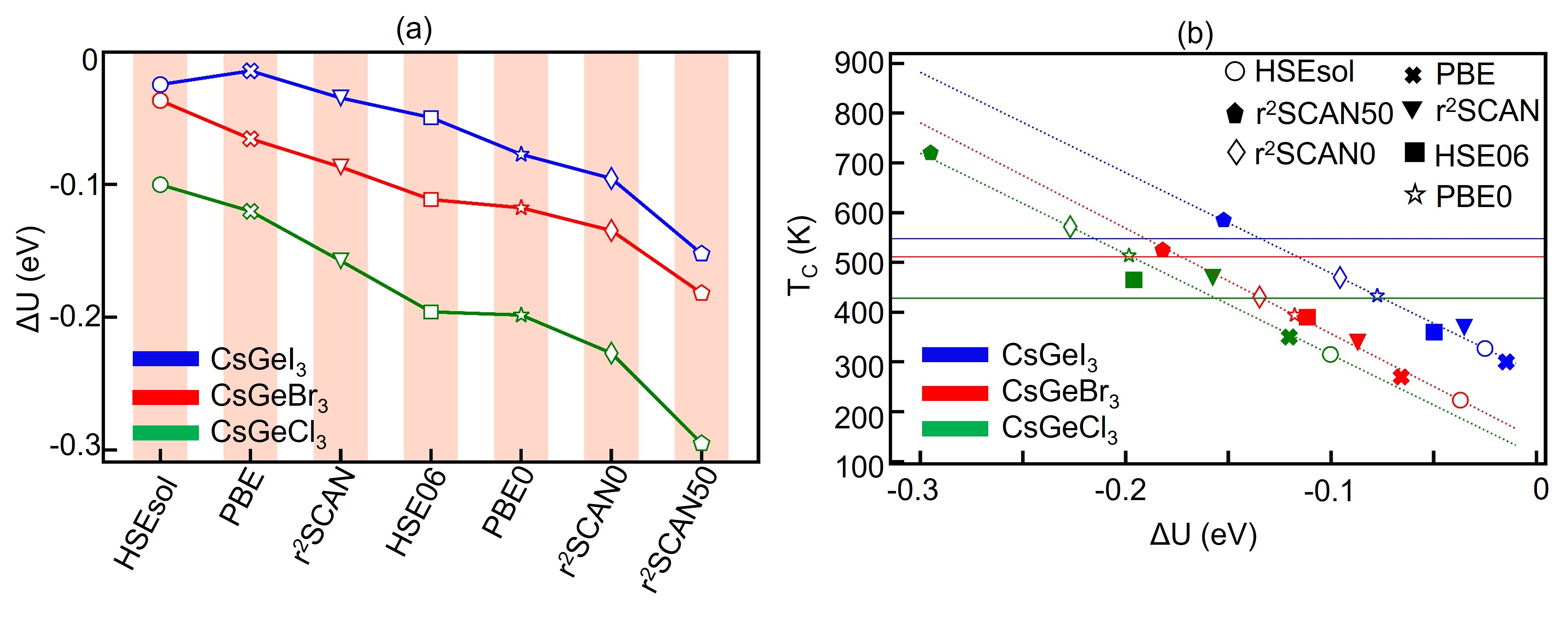}
\caption{ The energy difference between rhombohedral and cubic phases of  \cgx\ for different exchange-correlation functionals (a).  Curie temperature of  \cgx\ estimated from parametrization with different exchange-correlation functionals (b). Dashed lines give the linear fit to the data. Solid horizontal lines indicate experimental Curie temperatures.  }
\label{fig1}
\end{figure}

Figure~\ref{fig1}(a) gives  $\Delta U$ of  \cgx\ computed using different exchange-correlation functionals. We note that all functionals provide robust predictions for the relative ordering of $\Delta U$  for different materials in the \cgx~ family. Specifically, the stability of the rhombohedral phase increases as we go from I to Br and Cl. However, the actual energy difference is extremely sensitive to the choice of the exchange-correlation functional, the property noted in previous studies \cite{Maggie_PRM, XC-1, XC-2,XC-3, XC-4,XC-5}. In addition to the parametrization reported in Ref. \cite {Ravi_PRB} for \cgb, we have also parameterized the effective Hamiltonian for \cgx\, using HSE06 and r$^2$SCAN50 hybrid functionals. To compute the Curie temperature, we utilize the effective Hamiltonian within the framework of classical MD. The MD is used to simulate $NPT$ ensemble using Evans-Hoover thermostat\cite{rapaport2004art}. We use a simulation supercell of 30$\times$30$\times$30 unit cells of cubic perovskites with periodic boundary conditions applied along all three Cartesian directions. The supercell was  annealed  from  700~K to 10~K in steps of 10~K. For each temperature 300,000 MD steps were used, the first half of them was used for equilibration, while the last half is used to compute the thermal averages. Each MD step was 1 fs. The polarization at each step is computed from the average dipole moment of the simulation supercell. 
The transition temperatures predicted from such simulations using different effective Hamiltonian parametrizations are reported in Fig.~\ref{fig1}(b) as a function of $\Delta U$ (solid symbols). The data are fitted with a linear dependence, which allows for estimation of $T_C$  from other exchange correlation functionals. These estimations are given in Fig.~\ref{fig1}(b) by open symbols.   Comparison with experimental $T_C$ allows us to determine the best exchange-correlation functionals for the transition temperature predictions. They are r$^2$SCAN50 for X=Br, I and HSE06 for X=Cl, both hybrid functionals. 

\begin{table}[h]
\caption{Ground state properties for \cgx\ (X = Cl, Br and I) obtained from effective Hamiltonian and DFT  calculations and compared to experimental data from the literature. }
\centering
\begin{tabular}{|l|ccc|ccc|ccc|}
  \hline
  \multirow{2}{*}{} & \multicolumn{3}{c|}{CsGeCl$_3$} & \multicolumn{3}{c|}{CsGeBr$_3$} & \multicolumn{3}{c|}{CsGeI$_3$} \\ \cline{2-10}
  & H$_{eff}$ & DFT & Exp.\cite{CGI_exp_freq} & H$_{eff}$ & DFT & Exp. \cite{CGI_exp_freq}& H$_{eff}$ & DFT & Exp.\cite{CGI_exp_freq} \\ \hline
  P$_S$ ($\mu$C/cm$^2$) & 24.1 & 24.2 & - & 23.2 & 22.2 & 12-15 & 20.2 & 20.2 & $~$20 \\ \hline
  T$_C$ (K) & 460 & - & 428 & 525 & - & 511 & 590 & - & 548 \\ \hline
  $\omega_{E}$ (cm$^{-1}$)  & 178 & 209 & 237 & 176 & 168 & 160  & 139 & 147 & 151\\ \hline
  $\omega_{A1}$ (cm$^{-1}$) & 272 & 253 & 290 & 203 & 200 &  210 & 166 & 171 &  - \\  \hline
\end{tabular}
\label{propert}
\end{table}

We believe that the difficulty in reproducing the experimental data for any given functional stems from the difficulty of DFT to properly describe the lone pair \cite{PhysRevB.110.035160} in this compound, which is responsible for the stabilization of the rhombohedral phase. We note that our findings suggest that developing high-quality machine-learned potential for \cgx\ may be computationally prohibitive at this point, owing to the computational cost of hybrid functionals.  It is rather unexpected that among functionals tried we did not find one that works well for all three materials. This could be traced to the opposite trends in $\Delta U$ and $T_C$ for different materials in the family. Although, as already noted, $|\Delta U|$ increases from \cgi, to \cgb, to \cgc, suggesting a correlated increase in $T_C$, the experimental Curie temperature shows opposite trends, with $T_C$ increasing from \cgc, to \cgb, to \cgi.

Table \ref{propert} summarizes the predictions of the effective Hamiltonian compared to DFT and/or experimental data. We notice excellent agreement in  Curie temperatures between the effective Hamiltonian and the experiment.  Likewise, the agreement between DFT and effective Hamiltonian polarizations is excellent. We note that here we used an unusual strategy to parametrize the Born effective charge, $Z^*$, which is one of the effective Hamiltonian parameters. Typically, $Z^*$ is parametrized using the Born effective charges of individual ions computed in the cubic phase of the material. In our case, we opted to use  $Z^*=PV/u_0$, where $P$ and $V$ are the polarization and volume of the rhombohedral unit cell as obtained in DFT, while $u_0$ is the local mode that minimizes the Hamiltonian Eq.~(\ref{heff})\cite{PhysRevB.49.5828}.  In order to reproduce dynamical properties predicted from DFT we incorporate DFT frequencies of the $A_1$ mode into parametrization using the analytical expressions for the effective Hamiltonian  frequency: $\omega_{A_1}^2=-\frac{4k}{m}$, where $k$ is the quadratic term in the effective Hamiltonian \cite{Effective_H1} and $m$ is the local mode mass. Technically, we compute $m$ from DFT data for $\omega_{A_1}$ and $k$. Note that the analytical expression for $E$ mode can also be obtained  $\omega_E^2=\frac{1}{m}\frac{2k\gamma'}{3\alpha' + \gamma'}$, where $\alpha'$ and $\gamma'$ are the strain-renormalized fourth-order terms expansion coefficients for the effective Hamiltonian as defined in Ref. \onlinecite{PhysRevB.49.5828}.  To get soft mode frequencies we compute complex dielectric response, $\epsilon(\omega) = \epsilon'(\omega)+ i  \epsilon''(\omega)$, using the computational approach of Ref.~\onlinecite{magnon}. Here $\epsilon '$ and $\epsilon''$ are the real and imaginary parts of the response. At the lowest simulated temperature of 10 K the soft mode is underdamped and the location of the peak in $\epsilon''(\omega)$ can be used to estimate the mode frequency. The values obtained in this way are reported in Table \ref{propert}. The data demonstrated good agreement between DFT and effective Hamiltonian data and reasonable agreement with experimental data, where available. The parameters of the effective Hamiltonians for all three materials are given in Table \ref{T2_new}.

\begin{table}[h]
\caption{First-principles-based parameters for \cgx\, in atomic units using the notations of Ref.\onlinecite{Effective_H1} derived from hybrid functional as indicated in parenthesis next to the material.   }
\begin{tabular}{cccc}
\hline
Parameter & CsGeCl$_3$ (HSE06) & CsGeBr$_3$ (r$^2$SCAN50) &CsGeI$_3$ (r$^2$SCAN50) \\
\hline
\textit{a} ($\AA$) & 5.30    & 5.51 & 5.89  \\
\textit{Z$^{*}$} & 5.57 & 6.57 & 6.97 \\
\textit{$\epsilon_{\infty}$} &4.67 & 5.18 &6.80 \\
\textrm{$\kappa_2$}&-1.56 $\times$10$^{-3}$ & -2.91 $\times$10$^{-3}$ & -5.33 $\times$10$^{-3}$  \\
\textrm{$\alpha$}&6.44$\times$10$^{-2}$ & 8.00 $\times$10$^{-2}$& 7.72 $\times$10$^{-2}$ \\
\textrm{$\gamma$}& -10.65$\times$10$^{-2}$ & -11.97 $\times$10$^{-2}$& -11.22 $\times$10$^{-2}$ \\
\textit{j$_1$} & 0.82$\times$10$^{-3}$ & -2.77 $\times$10$^{-3}$ & -2.14 $\times$10$^{-3}$ \\
\textit{j$_2$} & -15.86 $\times$10$^{-4}$ & -25.13 $\times$10$^{-4}$ & -53.55 $\times$10$^{-4}$    \\
\textit{j$_3$} & 8.20$\times$10$^{-4}$ & 9.13 $\times$10$^{-4}$ & 6.51 $\times$10$^{-4}$  \\
\textit{j$_4$} & -4.30$\times$10$^{-4}$ & -4.84 $\times$10$^{-4}$ & -3.41 $\times$10$^{-4}$  \\
\textit{j$_5$} & -5.99 $\times$10$^{-4}$&  -4.24 $\times$10$^{-4}$ & -6.63 $\times$10$^{-4}$  \\
\textit{j$_6$} &  1.96 $\times$10$^{-4}$ & 2.22 $\times$10$^{-4}$ & 1.56 $\times$10$^{-4}$\\
\textit{j$_7$} & 0.00 & 0.00 & 0.00    \\
\textit{B$_{11}$} & 2.07& 2.33 & 2.29  \\
\textit{B$_{12}$} & 0.47& 0.45 & 0.40 \\
\textit{B$_{44}$} & 0.42& 0.45 &  0.51 \\
\textit{B$_{1xx}$} & -0.20& -0.41 & -0.18 \\
\textit{B$_{1yy}$} & -0.23& -0.22  & -0.20 \\
\textit{B$_{5xz}$} & -0.10&  -0.09  &  -0.11\\
\textit{m} & 36.21 & 88.56 & 109.95 \\
Eigenvector & (-0.02,-0.65,0.75,0.04,0.04) & (-0.00,-0.71,0.70,0.00,0.00) & (-0.0,0.78,-0.61,0.08,0.08) \\
\hline
\end{tabular}
\label{T2_new}
\end{table}

\section{Finite-temperature properties of C\MakeLowercase{s}G\MakeLowercase{e}X$_3$}
\subsection{Phase transition}

\begin{figure}[h]
\centering
\includegraphics[width=1\textwidth]{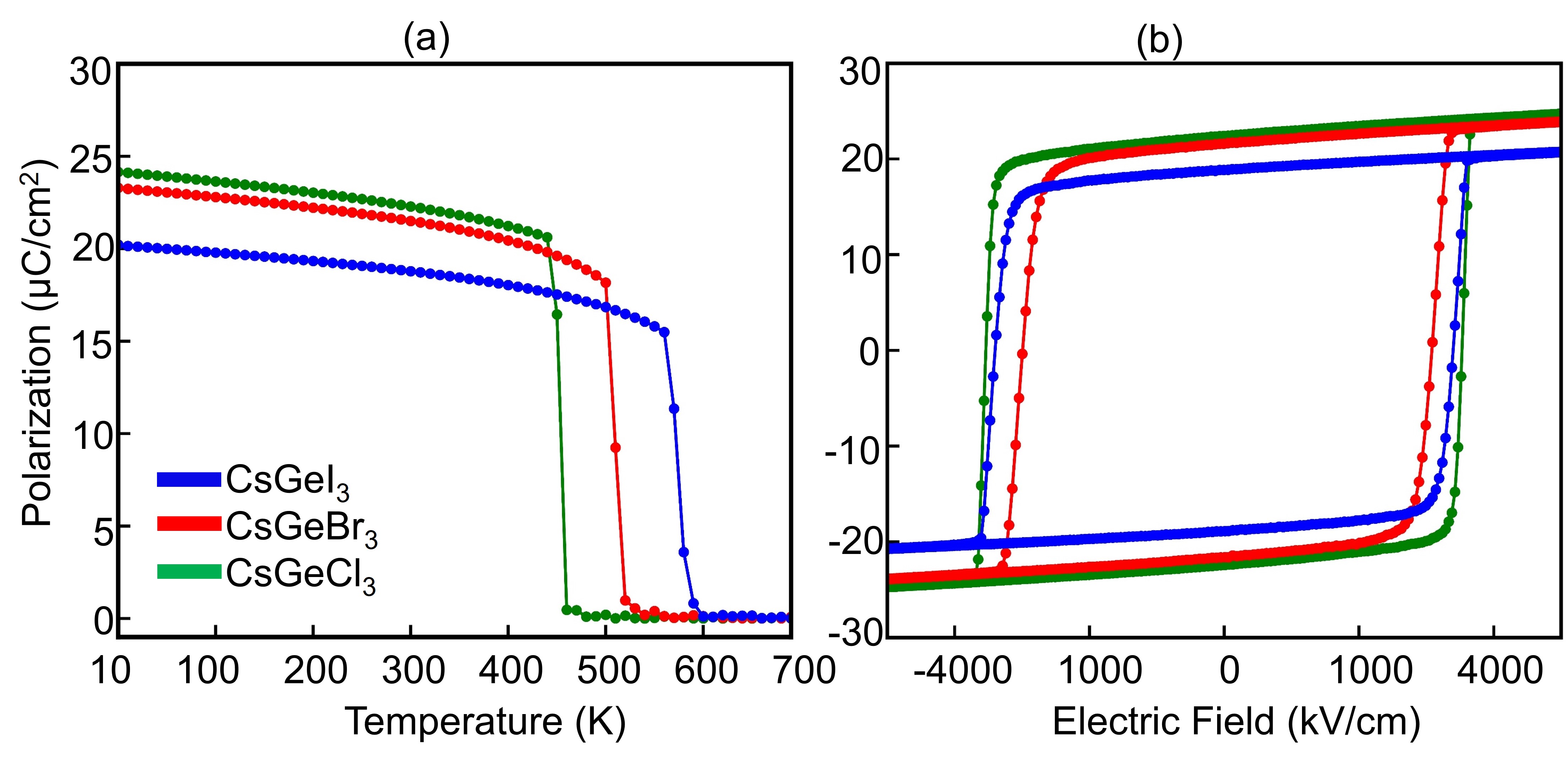}
\caption{Polarization as a function of temperature computed for \cgx\ (a). Hysteresis loops computed for \cgx\  (b). }
\label{fig2}
\end{figure}

\begin{figure}
\centering
\includegraphics[width=0.9\textwidth]{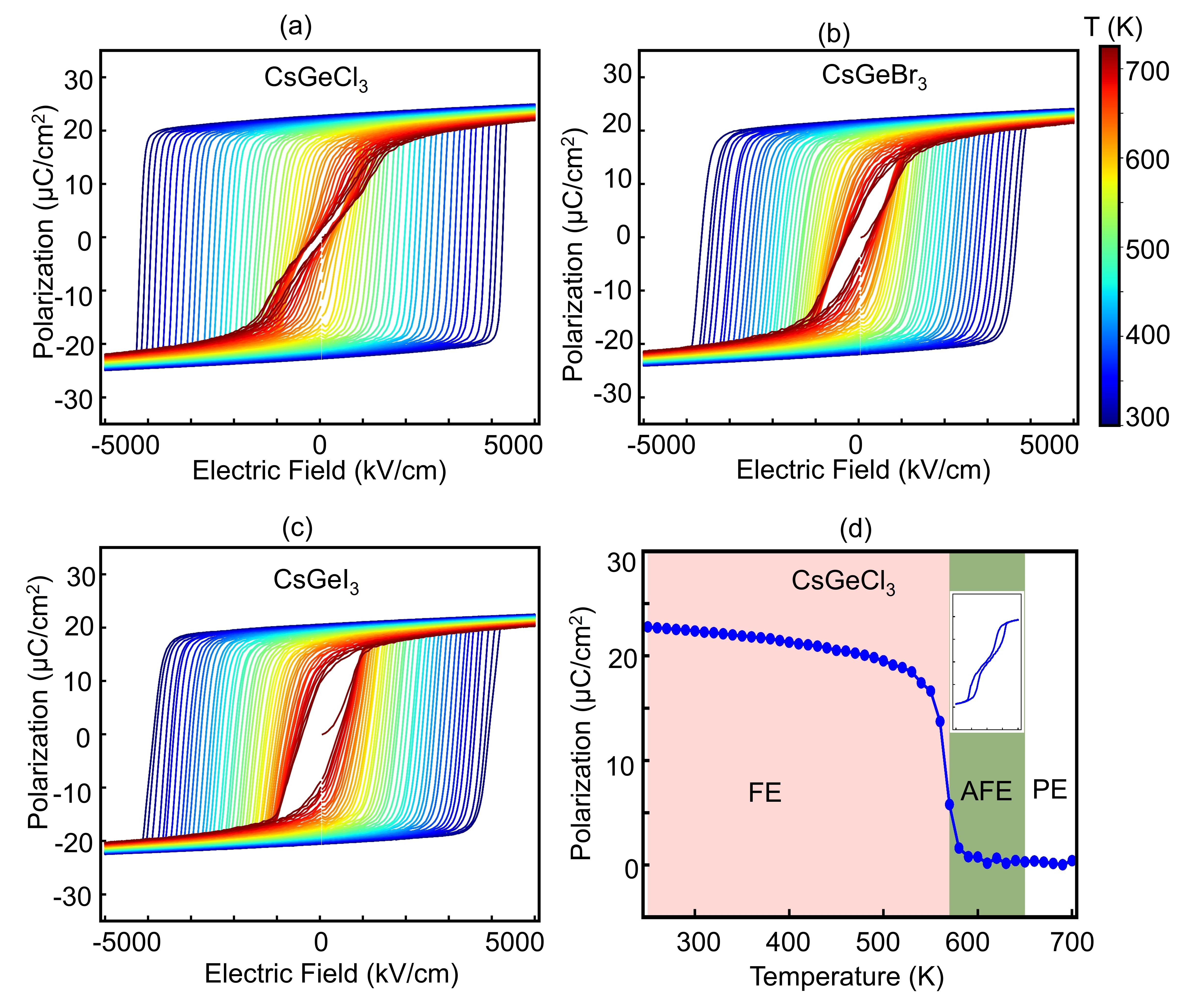}
\caption{   Hysteresis loops for \cgx\, computed for different temperatures (a)-(c). Polarization (E=0) as a function of temperature for \cgc\ from P-E loops (b). FE, AFE, and PE mark the regions associated with ferroelectric, antiferroelectric, and paraelectric hysteresis loops.  }
\label{fig3}
\end{figure}

We now use our parametrization to invetsigate finite-temperature properties  of \cgx. Figure~\ref{fig2}(a) shows the temperature evolution of polarization, which predicts a discontinuous change in polarization suggestive of a  mostly first-order type of phase transition. We find that the Curie temperature decreases as we go from \cgi\, to \cgb\, to \cgc, while polarization increases. This is rather unusual as typically the value of polarization correlates with $T_C$. We believe that it could be due to the fact that Curie temperatures are parametrized to experimental data, while polarizations are parametrized to DFT data. 
Figure~\ref{fig2}(b) shows hysteresis loops computed at 300 K and under an applied AC electric field of 5,000~kV/cm amplitude and 10~GHz frequency. We find that all three materials exhibit comparable responses at room temperature. We note the usual overestimation of coercive field in computations which is typically attributed   to the lack of defects, finite supercell size, and high frequency of the applied field.  We will address some of these below.   Since \cgx\, are semiconducting\cite{CGX_Ferro}, the presence of carriers in the conduction band destabilizes the ferroelectric phase\cite{carrier-1, carrier-2} and affects electric properties. Our predictions correspond to the insulating phases of the materials. 
Figure \ref{fig3} shows hysteresis loops computed for different temperatures. At low temperatures, all three materials exhibit square loops. Interestingly, \cgc\, exhibits double hysteresis loops corresponding to antiferroelectric behavior at temperatures 580-620~K, 100~K above the Curie point. The double hysteresis loops were traced to the electrostriction. If we set electrostrictive couplings to zero, then the double hysteresis loops disappear.  Indeed, occurrence of double hysteresis loops is possible in the vicinity of phase transition for ferroelectrics with first order phase transition\cite{lines2001principles}. Since electrostriction couples strain and ferroelectric mode, it promotes the phase transition's first-order character, which could lead to the double hysteresis loops if metastable phases have relatively long lifetimes.

%\begin{figure}
%\centering
%\includegraphics[width=0.9\textwidth]{Fig4.jpg}
%\caption{ (a)-(b) Probability density of local modes at below and near T$_C$. (c)-(d) Probability distribution of thermally averaged local modes below and near T$_C$.    }
%\label{fig3}
%\end{figure}

\subsection{Polarization reversal}
Next, we focus on understanding polarization reversal in \cgx\, and its temperature evolution. Firstly, we note that the nearly perfect square hysteresis loops reveal that the polarization reversal in our simulations is a nucleation-limited process, that is, it is associated with the formation of multiple tiny nuclei distributed relatively uniformly throughout the sample at some critical value of the electric field, the coercive field $ E_C$, in this particular case. Once nucleation starts, the polarization reversal proceeds very efficiently through  further nucleations and nuclei growth, which is responsible for the nearly vertical slope of $P(E_C)$. We will consider this process as an Arrhenius one. During polarization reversal, the system has to overcome the energy barrier $\Delta U = U_B-U_A$, where $U_B$ and $U_A$ are the energies associated with the maximum and (local) minimum in the double-well energy landscape of a ferroelectric. The probability of nucleus formation (through hopping over a barrier) is proportional to $e^{-\frac{(\Delta U)}{k_BT} }$. The rate of nuclei formation  is 
$\tau^{-1}=\tau^{-1}_0e^{-\frac{(\Delta U)}{k_BT}}$, where $\tau$ is the nucleation, or hopping,  time and $\tau_0$ is the nucleation, or hopping, time for $\Delta U\rightarrow$0. However, when  $\Delta U\rightarrow$0  the double well potential evolves into a single well one and the dynamics is governed by the soft mode frequency in the cubic phase. So we expect $\tau_0$  to be in the range of soft mode frequencies 1-10~THz. In the presence of the electric field $E$ opposite to the polarization direction the barrier is reduced by $EPV$, and we obtain $\tau=\tau_0e^{\frac{\Delta U-EPV}{k_BT}}$ for the nucleation time. In order to observe a hysteresis loop the nucleation, or switching, time should be less or equal to half a period for the electric field, $\mathcal{T}$, that is $\tau \leq \mathcal{T}/2$, or 
\begin{equation}
    \tau_0e^{\frac{\Delta U-EPV}{k_BT}} \leq \mathcal{T}/2. 
    \label{ineq}
\end{equation}
This inequality allows to derive an  analytical expression for the coercive field. For example, we can start with low values of $\mathcal{T}$ (high AC field frequency) so that no hysteresis loops can be obtained as the inequality is not satisfied. We can then increase $\mathcal{T}$ until the loops start forming. The value of $\mathcal{T}^{eq}$ at which the loops start forming turns inequality into an equality ( ``{\it eq}" superscript in $\mathcal{T}$) and  can be used to to compute the coercive field from $\tau_0e^{\frac{\Delta U-E_CPV}{k_BT}} = \mathcal{T}^{eq}/2$. We get 
\begin{equation}
    E_C=\frac{1}{PV}\left (\Delta U -k_BT \ln \frac{\mathcal{T}^{eq}}{2\tau_0}\right ) = E_{C,T=0}(1-\frac{T}{T_C^*})
    \label{eq: ecoer}
\end{equation}
where $E_{C,T=0}=\frac{\Delta U}{PV}$ and $T_C^*=\Delta U/k_B\ln \frac{\mathcal{T}^{eq}}{2\tau_0}$. We can see that $T_C^{*}$ is the temperature at which $E_C=0$ and may be different from the Curie temperature of the material. It is a function of the electric field period/frequency and provides analytical validation to the concept of ``dynamic" Curie temperature proposed in Ref.\onlinecite{PhysRevB.102.224109}. 
It is important to remember that Eq.~(\ref{eq: ecoer}) is valid in the regime where we can observe the  loops disappearing as switching time becomes greater than the electric field period. In addition to varying $\mathcal{T}$ to enter the regime of Eq.~(\ref{eq: ecoer}) validity, one can vary the temperature.  The inequality (\ref{ineq}) will be  violated  for low temperatures, since switching time increases as the temperature decreases. That means that keeping $\mathcal{T}$ constant we can loose hysteresis loops below certain temperature. This is exactly what happens in our simulations. We keep $\mathcal{T}$ constant but decrease simulation temperatures. For our chosen $\mathcal{T}$ the loops disappear below 250~K, signifying that the switching time is now greater than the electric field period. 

\begin{figure}
\centering
\includegraphics[width=1\textwidth]{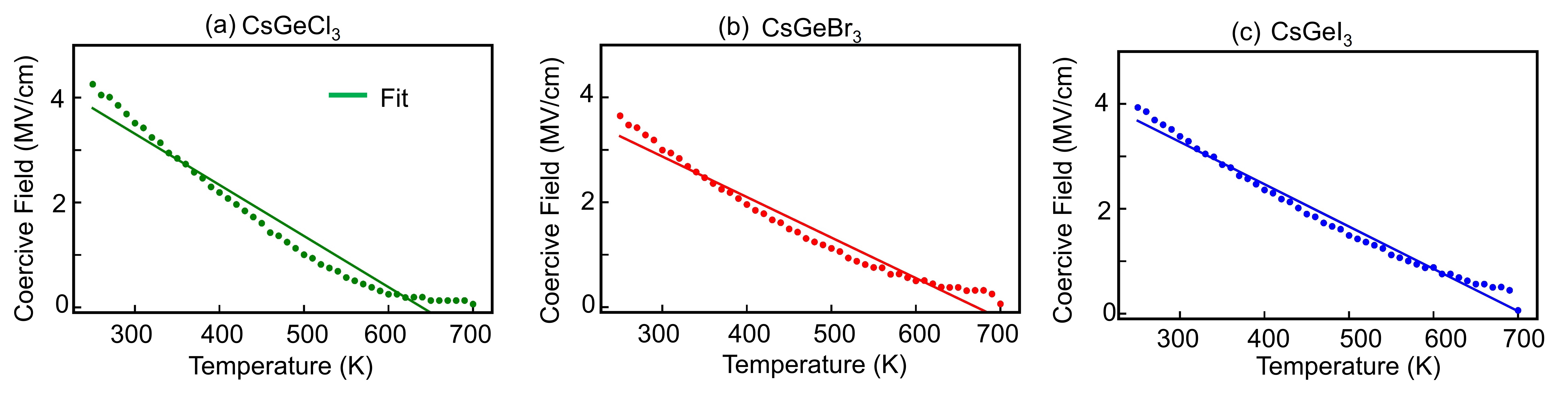}
\caption{  Coercive field as function of temperature Solid line gives linear fit to the data. }
\label{fig4}
\end{figure}

Let us now verify Eq.~(\ref{eq: ecoer}) by using our computational data. Figure \ref{fig4} shows the dependence of the coercive field computed from the hysteresis loops shown in Fig.~\ref{fig3} on temperature. We indeed find that $E_C(T)$ is very close to linear. Note, that the linear dependence has also been observed in a computational study on BaTiO$_3$ using machine-learned potentials \cite{FallettaStefano2025Udlo}.  We will discuss later the reasons behind the small nonlinearity. We next fit the data using Eq.~(\ref{eq: ecoer}). The fitting parameters $E_{C,T=0}$ and $T_C^*$ are used to obtain the barriers $\Delta U$ and intrinsic switching times $\tau_0$. The values are reported in Table~\ref{dynamics}. We first notice that the intrinsic switching times are in the range 0.4-0.6 ps (or 1.6-2.5 THz frequency), which indeed coincides with the soft mode frequency range, thus providing validation to out hypothesis about the value of $\tau_0$. Secondly, we can now gain an insight into polarization reversal by analyzing the barriers $\Delta U$ reported in the Table. To that end, let us consider polarization reversal that passes through the cubic phase, which becomes a transition state for such a process. The barrier for such a process can be estimated as the difference between the cubic and ground state energies. These values can be computed from both DFT simulations (see Fig.~\ref{fig1}(a))  and the analytical solution for the effective Hamiltonian model\cite{PhysRevB.49.5828}. They are reported in Table \ref{dynamics}.  We can see the $\Delta U$ obtained from the coercive field fits are significantly smaller than the their effective Hamiltonian and DFT counterparts. We believe that this is due to the fact that the actual transition state for the polarization reversal differs from the cubic phase. When we use cubic phase as the transition state we assume that the state is microscopically cubic\cite{PhysRevB.105.224108}, meaning that all the unit cells have zero polarization. However, in a large system the energy can be minimized (under constraint of zero polarization) through accessing microstates that are cubic only macroscopically, but not microscopically, or locally. One illustrative example of this is polarization reversal through domains. In such a case the macrostate indeed has zero   polarization, while microstate has nonzero local polarization.  Thus our findings provide the first direct confirmation that polarization reversal is associated with a transition state that is cubic only macroscopically even in defect-free samples.

\begin{table}[ht]
\caption{Polarization reversal barrier ($\Delta U$) and intrinsic switching times ($\tau_0$) obtained from effective Hamiltonian analytical solution (H$_{eff}$), DFT computations, and from the fitted values of $E_{C,T=0}$.}
\centering
\begin{tabular}{|l|c|c|c|c|c|c|c|c|c|}
  \hline
  \multirow{2}{*}{} & \multicolumn{3}{c|}{CsGeCl$_3$} & \multicolumn{3}{c|}{CsGeBr$_3$} & \multicolumn{3}{c|}{CsGeI$_3$} \\ \cline{2-10}
  & H$_{eff}$ & DFT & from $E_{C,T=0}$  & H$_{eff}$ & DFT & from $E_{C,T=0}$ & H$_{eff}$ & DFT &from $E_{C,T=0}$ \\ \hline
  $\Delta$U (meV/f.u.) & 196 & 174 & 136 & 182 & 165 & 120 & 152 & 201 & 139 \\ \hline
$\tau_0$ (ps) & - & - & 0.40 & - & - & 0.62 & - & - & 0.50 \\ \hline

\end{tabular}
\label{dynamics}
\end{table}

Now, that the Eq.~(\ref{eq: ecoer}) has been validated let's summarize the insights that it provides: (i) the nucleation limited polarization switching is an Arrhenius process that results in nearly linear dependence of the coercive field on the temperature at least in the regime where the switching time is comparable to the period of the AC electric field. This regime can be accessed by varying period of the electric field or temperature; (ii) It allows to deduce intrinsic switching time from $E_C(T)$ dependencies. This time is expected to coincide with the inverse of the soft mode frequency; (iii) it gives access to the actual polarization reversal barrier from the data on $E_C(T)$ which can be used to infer the nature of the transition state. For example, in our case it was found that the state is only macroscopically cubic; (iv) It establishes the dependence of the dynamical Curie temperature $T_C^*$ on the electric field period $\mathcal{T}$, which helps to understand why hysteresis loops computed at high AC field frequency may not be used to obtain Curie temperatures; (v) It reveals the actual reason  for  the coercive field ``overestimation" in simulations. To see that let us rewrite Eq.~(\ref{eq: ecoer}) as $E_C=\frac{1}{PV}\left (\Delta U -k_BT \ln \frac{\nu_0}{2\nu}\right )$, where $\nu_0$ is the soft mode frequency and $\nu$ is the electric field frequency. Let us further rewrite  $\ln \frac{\nu_0}{2\nu} \approx ln 10^n \approx n$, where $n$ is the order of magnitude difference between $\nu_0$ and $\nu$. So we have $E_C\approx \frac{1}{PV}\left (\Delta U -nk_BT \right ) $.  In our simulations $\nu_0/\nu \approx 10^{12}/10^{10}$ so that $n \approx 2$. Consequently, $E_C\approx \frac{1}{PV}\left (\Delta U -2 k_BT \right ) $. $\Delta U$ is on the order of hundreds meV (see Table~\ref{dynamics}),   temperature dependent term is on the order of tens of meV at room temperature, therefore, it does not make very large contribution to the $E_C$. However, if we could simulate MHz frequency, then  $E_C\approx \frac{1}{PV}\left (\Delta U -6 k_BT \right )$, which would considerably decrease the coercive field. This is consistent with the observations made in the computational study on BaTiO$_3$ \cite{FallettaStefano2025Udlo}.  Of course, one has to be careful whether the Eq.~\ref{eq: ecoer}  still remains somewhat valid in this regime, but nevertheless the argument reveals that the origin of discrepancy between the computational and experimental coercive field is the discrepancy between the timescale of simulations and experiment. In other words, simulations and experiments probe completely different regimes.

Let us now comment on the possible origin of the deviation of $E_C$ in computations (see Fig.~\ref{fig4}) from the linear dependence predicted by Eq.~(\ref{eq: ecoer}).
Strictly speaking, the model applies to the dipoles whose energy is exactly $U_A$, that is the energy minimum, as we do not average over energies $U$. The model can be improved by computing the thermal average  $<E_C>=\sum E_C(i)e^{-U_i/k_BT}$, where $E_C(i)$ is the coercive field associated with the dipoles whose state is described by the energy $U_i$. That way, the model will incorporate contributions from all the dipoles, not only the ones at the minimum. We believe this enhancement will reproduce the slight nonlinearity of $E_C(T)$.  However, in this case, we also need to know the entire  $U(P)$ which will make the model less general. Therefore, we limit the model to the dipoles at the minimum of energy to capture the basic physics.   

\subsection{Polar-based properties: dielectric susceptibility, pyroelectricity, energy storage }

\begin{figure}[h]
\centering
\includegraphics[width=1\textwidth]{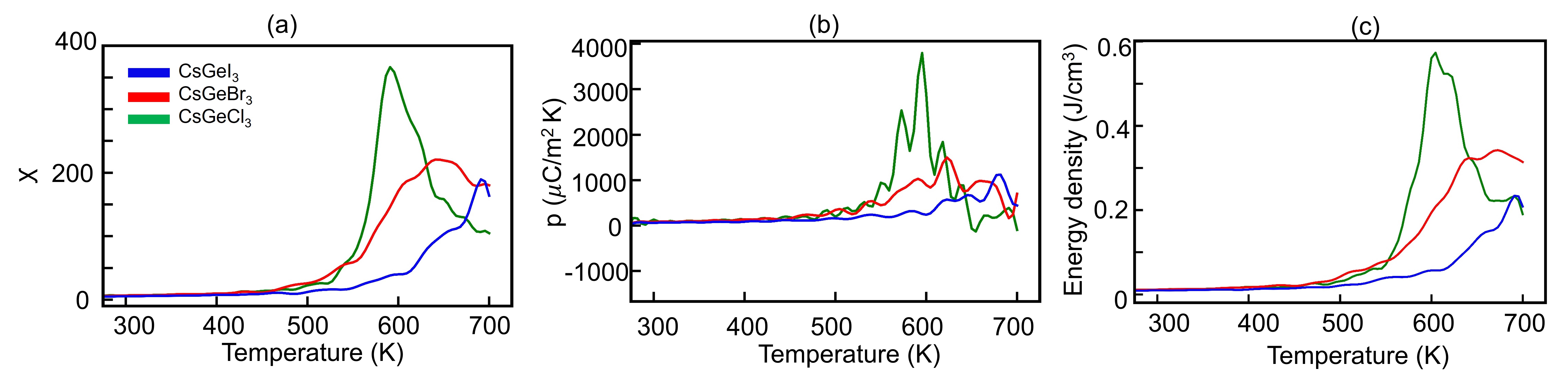}
\caption{  Susceptibility (a), pyroelectric coefficient (b) and  energy density (c) as a function of temperature computed for  \cgx.}
\label{fig5}
\end{figure}

We next use hysteresis loop data (Fig.~\ref{fig3}) to compute other finite-temperature properties, namely zero field dielectric susceptibility $\chi=\left (\frac{\partial P}{\partial E} \right )_{E=0}$, pyroelectric coefficient $p=\left (\frac{\partial P}{\partial T} \right )_{E=0}$ and energy storage density $u=\int_{P_{remn}}^{P_{max}} E dP $, where $P_{remn}$ and $P_{max}$ are remnant and maximum polarization achieved during capacitor charging, respectively.  We have used the computational data for $P(T, E)$ to train a Gaussian process regression model, which is then used to approximate  $\chi\approx \frac{P(E_2)-P(E_1)}{E_2-E_1}$ for different temperatures, where $E_2=-E_1=$50~kV/cm. Likewise $p(T)\approx \frac{P(T_2)-P(T_1)}{T_2-T_1}$, where $T_{2/1}=T\pm \Delta T $ with $\Delta T=$2~K.  The energy density is technically computed from the prediction of the model for $P(T, E)$ as $u=E_{max}P(E_{max})-\int_0^{E_{max}} PdE$, where $E_{max}=$200~kV/cm is the largest field applied to the capacitor. The data are given in Fig.~\ref{fig5}. We find that all three properties have the highest values for \cgc, followed by \cgb, and \cgi. In particular, we find that $\chi$ in \cgc\, can reach  362 in the vicinity of the transition temperature, which compares with SrTiO$_3$ and LiNbO$_3$, and is higher than HfO$_2$ \cite{LiNbO3, Hfo2}. The pyroelectric coefficient can reach up to 3800  $\mu$C/m$^2$K in \cgc\, near T$_C$.  For, \cgb\ and \cgi, the values are 1500 $\mu$C/m$^2$K and 1120 $\mu$C/m$^2$K, respectively. The room temperature values are 125 $\mu$C/m$^2$K, 80 $\mu$C/m$^2$K, and 62 $\mu$C/m$^2$K, respectively, which are comparable with standard oxide perovskite ferroelectrics\cite{Pyro_materials}. Energy density storage reaches 0.57 J/cm$^3$,  0.34 J/cm$^3$ and 0.23 J/cm$^3$ values in \cgc\, \cgb\ and \cgi\ which compares well  with titanium based oxider relaxor ferroelectrics \cite{Bunpang2024}

\section{Conclusions}

In summary, we have proposed a computational approach to parametrization of effective Hamiltonian for perovskites with lone-pair using the recently discovered family of ferroelectric semiconducting halide perovskites \cgx (X$=$Cl, Br, I). The strategy relies on using hybrid exchange correlation functionals to accurately describe the energy difference between the cubic and ground states of the ferroelectrics and pick the functional that provides the best agreement with experiment for the Curie temperature. The polarization and soft mode frequency can be fitted to DFT values. The strategy has been used to parametrize effective Hamiltonians for all three materials in the family. The Hamiltonians were used in the framework of classical MD simulations to predict finite-temperature properties of the materials, which include temperature evolution of spontaneous polarizations and ferroelectric hysteresis loops. Both were found to be comparable among the three materials. To understand the fundamentals of polarization reversal in ferroelectrics we developed a minimalistic model which predicts the dependence of coercive field on both temperature and AC field period in the regime where intrinsic switching time is comparable to the frequency of the electric field. The model allows for determining both intrinsic switching times and energy barrier for polarization reversal from the data on coercive field. It reveals that the fundamental origin of the notorious discrepancy between the coercive field in simulations and experiment is in the fact that they probe different regimes: high frequency in simulations vs low frequency in experiment. The finite-temperature data for dielectric susceptibility, pyroelectric coefficient, and energy storage density reveal that these halide perovskites should exhibit performance comparable to oxide ferroelectrics.  

{\it Acknowledgements. We thank Laurent Bellaiche for useful discussions. This work was supported by the U.S. Department of Energy, Office of Basic Energy Sciences, Division of Materials Sciences and Engineering under Grant No. DE-SC0005245. Computational support was provided by the National Energy Research Scientific Computing Centre (NERSC), a U.S. Department of Energy, Office of Science User Facility located at Lawrence Berkeley National Laboratory, operated under Contract No. DE-AC02-05CH11231 using NERSC Award No. BES-ERCAP-0025236.x. Other computational support was provided by the U.S. Department of Defence High Performance Computing Centre.

\bibliography{paper}

\end{document}